\documentclass[twocolumn, nofootinbib, showpacs]{revtex4}
\usepackage[brazil]{babel}
\usepackage[latin1]{inputenc}
\begin{document}

\title{Five-Dimensional Mechanics as the Starting Point for the
Magueijo-Smolin Doubly Special Relativity}

\author{B. F. Rizzuti } \email{brunorizzuti@ufam.edu.br}
\altaffiliation{On leave of absence from Instituto de Sa\'ude e
Biotecnologia, ISB, Universidade Federal do Amazonas, AM, Brazil.}

\author{A. A. Deriglazov}

\affiliation{Depto. de F\'isica, ICE, Universidade Federal de Juiz
de Fora, MG, Brazil}

\affiliation{Depto. de Matem\'atica, ICE, Universidade Federal de
Juiz de Fora, MG, Brazil}

\begin{abstract}
We discuss a way to obtain the doubly special relativity
kinematical rules (the deformed energy-momentum relation and the
nonlinear Lorentz transformations of momenta) starting from a
singular Lagrangian action of a particle with linearly realized
$SO(1,4)$ symmetry group. The deformed energy-momentum relation
appears in a special gauge of the model. The nonlinear
transformations of momenta arise from the requirement of
covariance of the chosen gauge.
\end{abstract}

 \maketitle {\bf Keywords:} Nonlinear
Lorentz transformations, Doubly special relativity models

\section{Introduction}
Various doubly special relativity (DSR) proposals have received a
great amount of attention in the last years \cite{gi, ms, jmls, kg,
kg1, cg, km, gr, aad, el, bfr, fsl, smig, obs, exp, ljm, sy, mig}.
They have been formulated on the base of nonlinear realizations of
the Lorentz group in four-dimensional space of the particle
momentum \cite{ms}. It can be achieved introducing, in addition to
the speed of light, one more observer independent
scale\footnote{The idea of another invariant scale in space-time,
besides $c$, is very old. Snyder has constructed a Lorentz
invariant space-time that admits an invariant length, in one of
the first attempts to avoid divergence problems \cite{s}.},
$\zeta$, the latter is associated to the Planck scale (for a
recent review, see \cite{sy}). In turn, the nonlinear realization
implies deformed energy-momentum dispersion relation of the form
\begin{equation}\label{i.1}
\eta_{\mu \nu}p^{\mu}p^{\nu}=-m^2c^2+ f(\zeta, p^0).
\end{equation}
It is supposed that in the limit $\zeta \rightarrow 0$ one
recovers the standard relation $p_{\mu}p^{\mu}=-m^2c^2$.

The attractive motivations for such kind of modification have been
discussed in the literature. There is
evidence on discreteness of space-time from non-perturbative
quantum gravity calculations \cite{rs}. The modified
energy-momentum relation implies corrections to the GZK cut-off
\cite{tk}, so DSR may be relevant for studying the threshold
anomalies in ultra-high-energy cosmic rays \cite{ct, jmls}.
Astrophysical data of gamma-ray bursts can be used for bounding
possible corrections to $p_{\mu}p^{\mu}=-m^2c^2$, see \cite{obs}.
In the recent work \cite{exp} it was suggested that experiments
with cold-atom-recoil may detect corrections to the
energy-momentum relations, and $f \neq 0$ in (\ref{i.1}) should be
interpreted as a quantum gravity effect.

In this work we discuss the initial Magueijo-Smolin (MS) proposal
\cite{ms}, which states that all inertial observers should agree
to take the deformed dispersion relation for the conserved
four-momentum of a particle
\begin{eqnarray}\label{101}
p^2=-m^2c^2(1+\zeta p^0)^2.
\end{eqnarray}
This is invariant under the following nonlinear transformations:
\begin{eqnarray}\label{102}
p'^{\mu}=\frac{\Lambda^\mu \, _\nu p^\nu}{1+\zeta(p^0 - \Lambda^0
\, _\nu p^\nu)}.
\end{eqnarray}
However, the list of kinematical rules of the model is not
complete, which raised a lively and controversial debate on the
status of DSR \cite{sy}. One of the problems consists of the
proper definition of total momentum for many particle system
\footnote{In \cite{aad} we observed that MS-type kinematics can be
related with linear realization of Lorentz group in
five-dimensional position space. On this base, an example of DSR
model free of the problem of total momentum has been
constructed.}. Due to non-linear form of the transformations,
ordinary sum of momenta does not transform as the constituents.
Different covariant composition rules proposed in the literature
lead to some astonishing effects, like the "soccer ball problem"
and the "rainbow geometry" \cite{ljm, gr}.

To understand these controversial properties, it would be
desirable to have in our disposal the relativistic particle model
formulated in the position space, which leads to DSR relations
(\ref {101}), (\ref{102}) in the momentum space. Despite a lot of
efforts \cite{el, cg, smig, bfr, fsl}, there appears to be no
wholly satisfactory solution of the problem to date. It is the aim
of the present work to construct the model that could be used as a
laboratory for simulations of the DSR kinematics.

Nonlinear realizations of the Lorentz group on the space of
physical dynamical variables often arise after fixation of a gauge
in a theory with the linearly realized Lorentz group on the
initial configuration space. Adopting this point of view, we study
in Section 2 a singular Lagrangian on five-dimensional position
space $x^A$, $A=(\mu, 4)$, $\mu=0, 1, 2, 3$, with linearly
realized $SO(1,4)$ group. To guarantee the right number of the
physical degrees of freedom, we need two first-class constraints.
The only $SO(1,4)$-invariant quadratic combinations of the
variables in our disposal are $p^2$, $xp$, $x^2$. We reject $x^2$
as it would lead to a curved space-time \footnote{There are
proposals considering the de Sitter as the underlying space for
DSR theories \cite{kg, kg1,cg, mig}.}. So, we look for the model
with the constraints $p^2 = 0$, $xp = 0$. They correspond to a
particle with unfixed four-momentum, and without five-dimensional
translation invariance. In Section 3 we show that the MS deformed
energy-momentum relation arises by fixing an appropriate gauge
(for one of the constraints), and the nonlinear transformation law
of momenta is dictated by covariance of the gauge. Section 4 is
left for conclusions.

\section{$SO(1,4)$\,-invariant mechanics}
The motion of a particle in the special relativity theory can be
described starting from the three-dimensional action $-mc^2\int\
dt\sqrt{1-(\frac{dx^i}{dx^0})^2}$. It implies the Hamiltonian
equations
\begin{equation}\label{103}
\frac{dx^i}{dx^0}=\frac{p^i}{\sqrt{\vec p^2+m^2c^2}}, \qquad
\frac{dp^i}{dx^0}=0.
\end{equation}
The problem here is that the Lorentz transformations,
$x'^{\mu}=\Lambda^\mu{}_\nu x^\nu$, act on the physical dynamical
variables $x^i(x^0)$ in a higher nonlinear way. To improve this,
we pass from the three-dimensional to four-dimensional formulation
introducing the parametric representation $x^i(\tau)$, $x^0(\tau)$
of the particle trajectory $x^i(x^0)$. Using the relation
$\frac{dx^i}{dx^0}=\frac{\dot x^i(\tau)}{\dot x^0(\tau)}$, the
action acquires the form $-mc\int d\tau\sqrt{-\eta_{\mu\nu}\dot
x^\mu\dot x^\nu}$. It is invariant under the local transformations
which are arbitrary reparametrizations of the trajectory,
$\tau\rightarrow\tau'(\tau)$. In turn, in the Hamiltonian
formulation the reparametrization invariance implies the Dirac
constraint which is precisely the energy-momentum relation
$(p^\mu)^2=-m^2c^2$. Presence of the constraint becomes evident if
we introduce an auxiliary variable $e(\tau)$ and rewrite the
action in the equivalent form, $S=\int d\tau(\frac{1}{2e}(\dot
x^\mu)^2-\frac{e}{2}m^2c^2)$. Then equation of motion for $e$
implies the Lagrangian counterpart of the energy-momentum
relation, $\frac{\delta S}{\delta e}\sim\dot x^2+e^2m^2c^2=0$.
Besides the constraint, the action implies the equations of motion
$\dot x^\mu=ep^\mu$, $\dot p^\mu=0$. The auxiliary variable $e$ is
not determined by these equations and enter into solution for
$x^\mu(\tau)$ as an arbitrary function. The ambiguity reflects the
freedom which we have in the choice of parametrization of the
particle trajectory. By construction, the ambiguity is removed
excluding the parameter $\tau$ from the final answers.
Equivalently, we can impose a gauge to rule out the ambiguity as
well as the extra variables. The most convenient gauge is $e=1$,
$x^0=p^0\tau$, as it leads directly to the equations (\ref{103})
for the physical variables.

In resume, to avoid a nonlinear realization of the Lorentz group
in special relativity, we elevate the dimension of space from 3 to
4. In the DSR case, the Lorentz transformations are non linear in
the four-dimensional space. So, by analogy with the previous case,
we start from a theory with the linearly realized group in
five-dimensional space. Consider the action
\begin{equation}\label{c.1}
S=\int d\tau\frac{m}{2}\eta_{AB}Dx^ADx^B,
\end{equation}
where $\eta_{AB}=(-1,+1,+1,+1,+1)$, $Dx^A$ stands for the
"covariant derivative", $Dx^A\equiv\dot x^{A}-g x^A$, and
$g(\tau)$ is an auxiliary variable. The action is invariant under
$SO(1,4)$ global symmetry transformations
\begin{equation}\label{c.1_1}
x^A \rightarrow x'^A=\Lambda^A{}_Bx^B.
\end{equation}
There is also the local symmetry with the parameter
$\gamma(\tau)$,
\begin{eqnarray}\label{c.1_2}
\tau \rightarrow \tau'(\tau); \, \frac{d
\tau'}{d\tau}=\gamma^2(\tau), \cr x^A(\tau) \rightarrow
x'^A(\tau')=\gamma(\tau)x^A(\tau), \cr g(\tau) \rightarrow
g'(\tau')=\frac{\dot \gamma(\tau)}{\gamma^3(\tau)}+
\frac{g(\tau)}{\gamma^2(\tau)}.
\end{eqnarray}
The transformation law for $g$ implies a simple transformation law
of the covariant derivative,
$Dx^A\rightarrow\frac{1}{\gamma}Dx^A$. Hence $g$ play the role of
the gauge field for the symmetry.

The presence of local symmetry indicates that the model presents
constraints in the Hamiltonian formulation. So we apply the Dirac
method \cite{pd} to analyze the action (\ref{c.1}). Introducing
the conjugate momenta, we find the expressions
\begin{equation}\label{c.2}
p_A=\frac{\partial L}{\partial \dot x^A}m(\dot x_A-g x_A), \quad
p_g=\frac{\partial L}{\partial \dot g}=0.
\end{equation}
Hence there is the primary constraint, $p_g=0$. The canonical
Hamiltonian $H_0$ and the complete Hamiltonian $H$ are given by
the expressions
\begin{equation}\label{c.4}
H_0=\frac{1}{2m}p_A^2+ gp_A x^A, \quad
H=H_0+\lambda p_g,
\end{equation}
where $\lambda$ is the Lagrange multiplier for the primary
constraint. The Poisson brackets are defined in the standard way,
and equations of motion follow directly
\begin{equation}\label{c.6}
\dot x^A=\frac{p^A}{m}+gx^A, \quad \dot p^A=-g p^A, \quad \dot
g=\lambda.
\end{equation}
From preservation in time of the primary constraint, $\dot p_g=0$,
we find the secondary constraint
\begin{equation}\label{c.8_1}
p_Ax^A=0.
\end{equation}
In turn, it implies the tertiary constraint
\begin{equation}\label{c.8_2}
p_A^2=0.
\end{equation}
The Dirac procedure stops on this stage, all the constraints
obtained belong to the first class.

Since we deal with a constrained theory, our first task is to
specify the physical-sector variables \cite{symme}. The initial
phase space is parameterized by 12 variables $x^A$, $p^B$, $g$,
$p_g$. Taking into account that each first-class constraint rules
out two variables, the number of phase-space physical variables is
$12-2\times 3=6$, as it should be for DSR-particle. We note that
Eq. (\ref{c.6}) does not determine the Lagrange multiplier
$\lambda$, which enters as an arbitrary function into solutions to
the equations of motion. According to the general theory \cite{pd,
gt, aad1}, variables with ambiguous dynamics do not represent the
observable quantities. For our case, all the initial variables
turn out to be ambiguous.

To construct the unambiguous variables, we note that the
quantities $\pi^\mu=\frac{p^\mu}{p^4}$,
$y^\mu=\frac{x^\mu}{x^4}$,
obey $\dot\pi^\mu=0$, $\dot{y}^\mu=\frac{e}{m}(\pi^\mu-y^\mu)$,
where $e\equiv\frac{p^4}{x^4}$. Since these equations resemble
those for a spinless relativistic particle, the remaining
ambiguity due to $e$ has the well-known interpretation, being
related with reparametrization invariance of the theory. In
accordance with this, we can assume that $y^\mu(\tau)$ represent
the parametric equations of the trajectory $y^i(t)$. The
reparametrization-invariant variable $y^i(t)$ has deterministic
evolution $\frac{dy^i}{dt}=c\frac{\pi^i-y^i} {\pi^0-y^0}$.

We can also look for the gauge-invariant combinations on the
phase-space. The well known remarkable property of the Hamiltonian
formalism is that there are the phase-space coordinates for which the Hamiltonian
vanishes \cite{aad1}. In these coordinates trajectories look like
the straight lines. For the case, the unambiguous variables with
this property are $\pi^\mu$, $\tilde x^\mu\equiv y^\mu-\pi^\mu$.

\section{DSR gauge}
In this Section we reproduce the MS DSR kinematics starting from
the $SO(1,4)$ model. First, we obtain the MS dispersion
relation (\ref{101})
imposing a particular gauge in our model. Generally, neither the
global nor the local symmetries survive separately in the gauge
fixed version. But we can look for their combination that does not
spoil the gauge condition. Following this line, we arrive at the
MS transformation law of the momenta (\ref{102}).

According the Dirac algorithm, each first class constraint must be
accompanied by some gauge condition of the form $h(x, p)=0$, where
the function $h$ must be chosen such that the system formed by
constraints and gauges is second class. The constraints and the
gauges then can be used to represent part of the phase space
variables through other. Equations of motion for the remaining
variables are obtained by substituting the constraints and gauges
into the equations already found.

Let us choose the gauge $g=0$ for the constraint $p_g=0$. This
gauge fixes the local symmetry, as it should be,
\begin{equation}\label{c.9}
g'=\frac{\dot
\gamma}{\gamma^3}+\frac{g}{\gamma^2}\Big|_{g=0}\Rightarrow \dot
\gamma=0.
\end{equation}

We are, then, left with two constraints. To obtain a deformed
dispersion relation, we impose the gauge $p^4=mc h(\zeta, p^0)$
for the constraint $p_A x^A=0$. Using this expression in the
constraint (\ref{c.8_2}), we obtain
\begin{equation}\label{a.1}
p_{\mu}p^{\mu}=-m^2c^2 h^2(\zeta, p^0).
\end{equation}
We wrote the function $h$ depending on the arguments $\zeta$ and
$p^0$ but one is free to chose the particular dispersion relation
he wants. We point out that the scale $\zeta$ is
gauge-noninvariant notion in this model \footnote{It is worth
noting that a gauge-fixed formulation, considered irrespectively to the initial one,
generally has the physical sector different from those of the
initial theory \cite{gt}.}.

We now turn to the induced nonlinear Lorentz transformation of
momenta. Under the symmetries (\ref{c.1_1}), (\ref{c.1_2}), the
conjugated momentum $p^A=mDx^A$ transforms as
\begin{equation}\label{c.11}
p^A \rightarrow p'^A=\frac{1}{\gamma}\Lambda^A{}_B p^B.
\end{equation}
For $SO(1,3)$\,-subgroup \footnote{We discuss only the induced
Lorentz transformations. The remaining transformations are boosts
in the fifth dimension. In the gauge-fixed formulation they
produce the nonlinear transformations which play the role of
four-dimensional translations.}
\begin{eqnarray}\label{c.12}
\Lambda^A{}_B = \left( \begin{array}{cccccc}
\Lambda^{\mu}{}_{\nu} & 0  \\
0 & 1  \\
\end{array} \right),
\end{eqnarray}
we have
\begin{eqnarray}\label{c.13}
p^{\mu} \rightarrow
p'^{\mu}=\frac{1}{\gamma}\Lambda^{\mu}{}_{\nu}p^{\nu}, \, p^4
\rightarrow p'^4 =
\frac{1}{\gamma}\Lambda^4{}_Ap^A=\frac{1}{\gamma} p^4.
\end{eqnarray}
Now, as it often happens in gauge theories, global symmetry of the
gauge-fixed formulation is a combination of the initial
global symmetry and local symmetry with specially chosen parameter
$\gamma$. Since the gauge $p^4=mch(\zeta, p^0)$ is not preserved
by the transformations (\ref{c.1_1}) and (\ref{c.1_2}) separately,
one is forced to search for their combination, (\ref{c.13}), which
preserves the gauge. Imposing the covariance of the gauge
\begin{equation}\label{a.2}
p^4=mch(\zeta, p^0) \Leftrightarrow p'^4=mch(\zeta, p'^0),
\end{equation}
we obtain the equation for determining $\gamma$
\begin{equation}\label{a.3}
h(\zeta, p^0)=\gamma h(\zeta,
\frac{1}{\gamma}\Lambda^0{}_{\mu}p^{\mu}).
\end{equation}
In the gauge $g=0$, we have $p_A=\mbox{const.}$ on-shell, so Eq.
(\ref{a.3}) is consistent with (\ref{c.9}). Eqs. (\ref{c.13}) with
this $\gamma$ provides a non linear realization of the Lorentz
group which leaves invariant the deformed energy-momentum relation
(\ref{a.1}).

Let us specify all this for MS model. If we fix the gauge
$p^4=mc(1+\zeta p^0)$, the constraint $p_A^2=0$ acquires the form
of MS dispersion relation (\ref{101}). Enforcing covariance of the
gauge, the equation (\ref{a.3}) for determining $\gamma$ reads
\begin{equation}\label{a.4}
1+ \zeta p^0= \gamma (1+\zeta
\frac{1}{\gamma}\Lambda^0{}_{\mu}p^{\mu}).
\end{equation}
So, $\gamma$ is given by
\begin{equation}\label{c.16}
\gamma=1+\zeta(p^0-\Lambda^0{}_{\mu}p^{\mu}).
\end{equation}
Using this $\gamma$ in Eq. (\ref{c.13}), we see that the momenta
$p^{\mu}$ transform according to Eq. (\ref{102}).

\section{Concluding Remarks}
We have constructed an example of the relativistic particle model
(\ref{c.1}) on five-dimensional flat space-time with linearly
realized $SO(1,4)$ group of global symmetries and without the
five-dimensional translation invariance. Due to the local symmetry
presented in the action, the number of physical degrees of freedom
of the model is the same as for the particle of special relativity
theory. We have applied the model to simulate kinematics of the
Magueijo-Smolin doubly-special-relativity proposal. It was done by
an appropriate fixation of a gauge for the constraint
(\ref{c.8_1}), that leads to the MS deformed dispersion relation
(\ref{101}). The nonlinear transformation law of momenta
(\ref{102}) was found from the requirement of covariance of the
gauge-fixed version.

We finish with the comment on a transformation law for the spatial
coordinates. Using the parameter $\gamma$ obtained in Eq.
(\ref{c.16}), the transformation of the configuration-space
coordinates can be found from (\ref{c.1_1}) and (\ref{c.1_2})
\begin{equation}\label{c.18}
x^{\mu} \rightarrow
x'^{\mu}=[1+\zeta(p^0-\Lambda^0{}_{\mu}p^{\mu})]\Lambda^{\mu}{}_{\nu}x^{\nu},
\end{equation}
\begin{equation}\label{c.19}
x^4 \rightarrow x'^4=[1+\zeta(p^0-\Lambda^0{}_{\mu}p^{\mu})]x^4.
\end{equation}
The component $x^4$ is affected only by a scale factor. The
coordinates $x^{\mu}$ transform as usually happens in DSR
theories: we have a transformation law that is energy-momentum
dependent. These transformations were obtained in the work
\cite{km} from the requirement that the free field defined on DSR
space (\ref{101}) should have plane-wave solutions of the form
$\phi \sim Ae^{-ip_{\mu}x^{\mu}}$, then the contraction
$p_{\mu}x^{\mu}$ must remain linear in any frame. We point out
that it turns out to be true in our model
\begin{equation}\label{c.20}
\eta_{\mu \nu}p'^{\mu} x'^{\nu}=\eta_{\mu
\nu}(\frac{1}{\gamma}\Lambda^{\mu}{}_{\alpha}p^{\alpha})( \gamma
\Lambda^{\nu}{}_{\beta}  x^{\beta})=\eta_{\alpha \beta}p^{\alpha}
x^{\beta}.
\end{equation}
Eq. (\ref{c.18}) leads also to the energy-dependent metric of the
position space \cite{gr}. There are some attempts to interpret
$p^0$ in this case, see \cite{km, gr}.

\section*{Acknowledgments}
B. F. R. would like to thank the Brazilian foundation FAPEAM -
Programa RH Interiorização - for financial support. A. A. D.
acknowledges support from the Brazilian foundation FAPEMIG.

\end{document}